\documentstyle[11pt,twoside,jltp]{article}
\input epsf
\epsfverbosetrue

\title{Quantum Measurement Operations with Nanomagnets and SQUIDs}

\author{Martin Dub\'e\address{Helsinki Institute of Physics, P.O. Box 9
(Siltavuorenpenger 20 C), Helsinki FIN-00014, Finland} and Philip C. E.
Stamp\address{Department of Physics and Astronomy, and Canadian
Institute for Advanced Research, 6224 Agricultural Rd,
University of British-Columbia, Vancouver BC, Canada V6T 1Z1}}

\runninghead{M. Dub\'e and P. C. E. Stamp}{The Measurement Operation}
       
\begin{document}

\begin{abstract}
The low-energy behaviour of 2 coupled nanomagnets or 2
coupled SQUIDs, interacting with their environment, 
can be described by the model of a ``Pair or Interacting Spins
Coupled to an Environmental Sea'' (PISCES).
These physical systems can then be used for a measurement operation
in which system, apparatus and environment are all treated quantum
mechanically. We design a ``Bell/Coleman-Hepp'' measuring system,
and show that in principle one may design a situation in which
quantum interference between system and apparatus can upset the usual
measurement operation.

\end{abstract}

\maketitle


\section{INTRODUCTION}

The question of whether quantum mechanics applies at the mesoscopic or
macroscopic levels has generated an enormous amount of work in recent 
years. As emphasised by Leggett \cite{AJL}, theoretical discussions of  
this topic should include realistic physical models, 
with the ultimate goal of testing the predictions of quantum mechanics 
against experimental results. 
In particular, it is crucial to include all
couplings between the quantum systems and their environment (which according 
to Quantum theory, is itself also quantum-mechanical). 

In this context, it is surprising that no discussion of a Quantum
Measurement Operation ever considered both system and apparatus, both
coupled to some external environment, with all three described in a 
fully quantum mechanical way. Apart from the practical interest of such a
discussion (see below), it is also essential if one is to 
examine the cornerstone of the Copenhagen interpretation, viz., that quantum
states can only be defined with reference to some measuring system. 

We consider here a fully quantum-mechanical description of a 
measurement scheme using
systems that have already shown evidence of macroscopic quantum
behaviour, ie., SQUIDs and nanomagnets. The low-energy dynamics can
be reduced to the ``PISCES'' model, the dynamics of which was recently 
elucidated \cite{PIS1}. 

\section{Models of Quantum Measurements}

A quantum measurement typically involves some ``measuring
coordinate'' (usually macroscopic), and a quantum system 
which is being measured.  Both are 
coupled to their environment.  The simplest example is where the  
quantum ``system'' is stable in one of 2 interesting quantum 
states, and the apparatus likewise.  In this case the apparatus
works as a measuring device if its final state is
correlated with the initial state of the system.

One common such model, 
the ``Bell/Coleman-Hepp'' model \cite{hepp}, has Hamiltonian 
\begin{equation}
H^{(A)}_{o} (\vec{\tau}) \ + \ H^{(s)}_{o} (\vec{\sigma}) \ + \
\frac{1}{2}
K \ \hat{\tau}_x (1-\hat{\sigma}_z)
\label{1}
\end{equation}
where $\vec{\tau}$ is a Pauli vector describing the apparatus, and 
$\vec{\sigma}$ a Pauli vector for the system.  
The coupling $K = \pi$, so that if $\vec{\sigma}$ is in initial state 
$|\uparrow >$, then $\vec{\tau}$ is unaffected, whereas if
$\vec{\sigma}$ is initially $|\downarrow >$, then 
$\vec{tau}$ flips (with no change in $\vec{\sigma}$).  
We thus have the classic ``ideal measurement''
scheme, for which
\begin{eqnarray}
|\Uparrow> |\uparrow> \longrightarrow |\Uparrow> |\uparrow> \nonumber
\\
|\Uparrow> |\downarrow> \longrightarrow |\Downarrow> |\downarrow>  \;
\label{2}
\end{eqnarray}
ie., the final state of the apparatus is uniquely correlated with the
initial state of the system.

Such a scheme can in principle be obtained starting with a
large variety of initial ``microscopic'' high-energy Hamiltonians.  
Consider, eg., a Hamiltonian
\begin{eqnarray}
H = \frac{1}{2}(M_A q^{2}_{A} + M_s q^{2}_{s}) + V_A(q_A) + V_s (q_s) -
\xi_Aq_A
\nonumber \\  -  K_oq_sq_A + \frac{1}{2} \sum^{N}_{k=1}
m_k(\dot{x}^{2}_{k} +
\omega^{2}_{k}x^{2}_{k})
+ \sum^{N}_{k=1}(c^{(s)}_{k}q_s + c^{(A)}_k q_A)x_k ~\;
\label{3}
\end{eqnarray}
where $q_A$ and $q_S$ are the ``coordinate'' of the apparatus and system
respectively, $V_s(q_s)$ and $V_A(q_A)$
describe symmetric 2-well systems (with minima at $q_s = \pm 1,
q_A = \pm 1$) and the coupling $-K_oq_Aq_s$ is 
produced by
expanding the interaction in powers of $q_A$ and $q_s$. 
A ``bias field'' $\xi_A$ also acts on the apparatus. 
The environment is composed of delocalised harmonic oscillators $x_k$
modes, with linear coupling to the macroscopic systems. 

Below a ``UV cut-off'' energy $\Omega_0$, only the 2 lowest levels of each
system will be important;
thus if $kT, K_0, \xi_A \ll \Omega_0$, it is straightforward to truncate
Eq. (\ref{3}) to the low-energy Hamiltonian
\begin{equation}
H = H_0 ( \tau , \sigma) +
H_{osc} (\{ x_k \} ) + H_{int} (\tau ,
\sigma ; \{ x_k \} )
\label{htotal}
\end{equation}
where $H_{osc}$ and $H_{int}$ are the same as in Eq.
(\ref{3}) (with $q_A$ and $q_s$ replaced by $\tau_z$ and
$\sigma_z$, and with the restriction that all 
$\omega_k \ll \Omega_0$); and  
\begin{equation}
H_0 = \Delta_A \tau_x + \Delta_s \sigma_x - \xi_A
\hat{\tau}_z - {\cal K} \tau_z \sigma_z
\end{equation}
Here $\Delta_A$ and $\Delta_s$ are calculable tunneling matrix elements 
and ${\cal K}$ is a renormalised interaction, 
produced by integrating out high energy environmental 
modes ($\omega_k > \Omega_0$) 

We refer to Hamiltonians like (\ref{htotal}) as PISCES Hamiltonians
(PISCES being an acronym for ``Pair or Interacting Spins Coupled to an
Environmental Sea''). We have recently solved the  
dynamics of this model \cite{PIS1}. Here we shall (a) describe under what
conditions it behaves like the ideal measuring scheme (\ref{2}), 
and (b) discuss how
microscopic theory for SQUIDs or nanomagnets, operating in the quantum
regime, shows they are described by (\ref{htotal}). This then allows 
us to see how one may experimentally probe the quantum measurement 
operation.

\section{Nanomagnets and SQUIDs}

Consider as an example 2 conducting nanomagnets imbedded in a metallic 
matrix- they interact via dipolar interactions and also via the conduction 
electron bath. For this problem (of great importance in disordered magnets,
metallic glasses, and ``colossal magnetoresistance'' systems), the interaction 
with nuclear spins is not important (unlike for insulating nanomagnets!).
In the quantum regime, each nanomagnet behaves \cite{stamp} 
like a ``spin-boson'' system
\cite{legRMP}, with an Ohmic coupling to the bath of dimensionless strength
$\alpha_{\beta} \sim g^2 S_{\beta}^{4/3}$, where $\beta = 1,2$ 
labels each nanomagnet,
$g = JN(0)$ is the dimensionless 
exchange coupling at each electronic spin site in the nanomagnet, and $S$
is the spin quantum number of the ``giant spin'' produced when the electronic 
spins lock together at low $T$. Since typically $g \sim 0.05-0.2$ for metals, 
we get 
very strong coupling ($\alpha \gg 1$), even for very small metallic 
nanomagnets; however one can also vary $N(0)$, going even to semimetals, to
produce very small $\alpha$. In
isolation, if $\alpha \gg 1$, 
these ``giant Kondo'' spins are frozen by the electron bath until 
either $kT, \xi > \Omega_0$, where $\xi = g_{\mu_B}SH_o$ is the bias 
in an external field along the easy axis, and $\Omega_0$ is roughly the 
single ion anisotropy ($\sim 1K$), and independent of $S$.
   Truncation of the pair of nanomagnets to the quantum regime produces a 
PISCES Hamiltonian like eq. (\ref{htotal}), with a renormalised interaction
${\cal K}\tau^z \sigma^z = 
[V_{dip}({\bf R}) + {\cal J}(R)] \tau^z \sigma^z$, where ${\bf R}$ is the 
radius vector between the nanomagnet centres, and $V_{dip}$ is the dipolar 
interaction; transverse spin-spin couplings are irrelevant. The 
electron-mediated interaction ${\cal J}(R)$ depends on the shape and surface
properties of each nanomagnet (for analytic calculations see \cite{2FM}), 
but quite generally
\begin{equation}
{\cal J}(R) \sim \frac {\sqrt{\alpha_1 \alpha_2}}{(2 k_F R)^3} {\cal E}_F
\equiv \alpha_{12}(R) {\cal E}_F
\end{equation}
where ${\cal E}_F$ is the 
Fermi energy (ie., the bandwidth, in this simple calculation), and $k_F$ the 
Fermi wave-vector.

As a quite different example consider 2 interacting SQUIDs. The results
are intuitively obvious; we get a PISCES model, with  
coupling described by a mutual inductance 
and capacitance, and possibly a dissipative coupling through the
circuitry (depending on the experimental set-up); we give details elsewhere. 
One can also consider a SQUID ring coupled to a nanomagnet, with one or other 
behaving as a measuring device.

\section{Dynamics of the PISCES Model}

Let us briefly review the
dynamics \cite{PIS1} of the PISCES model- controlled
by the direct interaction ${\cal K}$, the temperature $T$, two  
renormalised tunneling matrix elements $\Delta_{\beta}$, and 
3 dissipative couplings, which we will assume here to be 
Ohmic, with values $\alpha_{\beta}$ and $\alpha_{12}(R)$. There are  
4 main dynamical regimes (as shown in Fig. (1) for the symmetric case 
$\alpha_1 = \alpha_2 = \alpha$, and $\Delta_1 = \Delta_2 = \Delta$), viz: 

(i) The ``Locked Phase'' (${\cal K} \gg T, \Delta_{\beta}/\alpha_{\beta}$):
The 2 spins lock together, in either ferro- or antiferromagnetic states 
depending on the sign of ${\cal K}$, and 
behave like a single spin-boson system, with tunneling matrix element
$\Delta_{c} = \Delta_{1} \Delta_{2} / | {\cal K}|$ and coupling 
$\alpha_{c}(R) = \alpha_{1}+\alpha_{2} \pm \alpha_{12}(R)$ 
to the oscillator bath. 

(ii) The ``Mutual Coherence'' phase ($\Delta_{\beta}/\alpha_{\beta} \gg T \gg
{\cal K}$; ${\cal K} > \Delta_{\beta}$): The thermal energy 
$\gg {\cal K}(R)$, but if the dissipative couplings 
$\alpha_{\beta}$'s are small ($\alpha_{\beta} \ll 1$), 
the energy scale $\Delta_{\beta}^{*}/\alpha_{\beta}$ can  
dominate even if $\Delta_{\beta} < {\cal K}$. The 2 systems are then in
a superposition of quantum states, and the dynamics is
composed of damped ``beat''-like oscillations, at 3 different frequencies. 

(iii) The ``Correlated Relaxation'' or High-T phase ($T \gg \Delta_{\beta}/
\alpha_{\beta}, {\cal K}$). Each spin 
relaxes incoherently, but  
in the time-dependent bias
generated by the other- there are 3 characteristic relaxation times.
These oscillations simply reflect the fact that the system is in a
superposition of quantum states. 

(iv) The rather boring ``perturbative regime''
(${\cal K} \ll \Delta_{\beta}^{*}$), in which the total coupling 
only weakly affects the single spin-boson behaviour 

An important feature of the PISCES model is that as soon as 
${\cal K} > \Delta_{\beta}$, the single spin-boson model results 
\cite{AJL} are invalid, and decoherence effects are massively amplified. 
Coherent oscillations can only exist in 
the Mutual Coherence regime. 

\begin{figure}[t]
\epsfysize=4.5in
\epsfbox{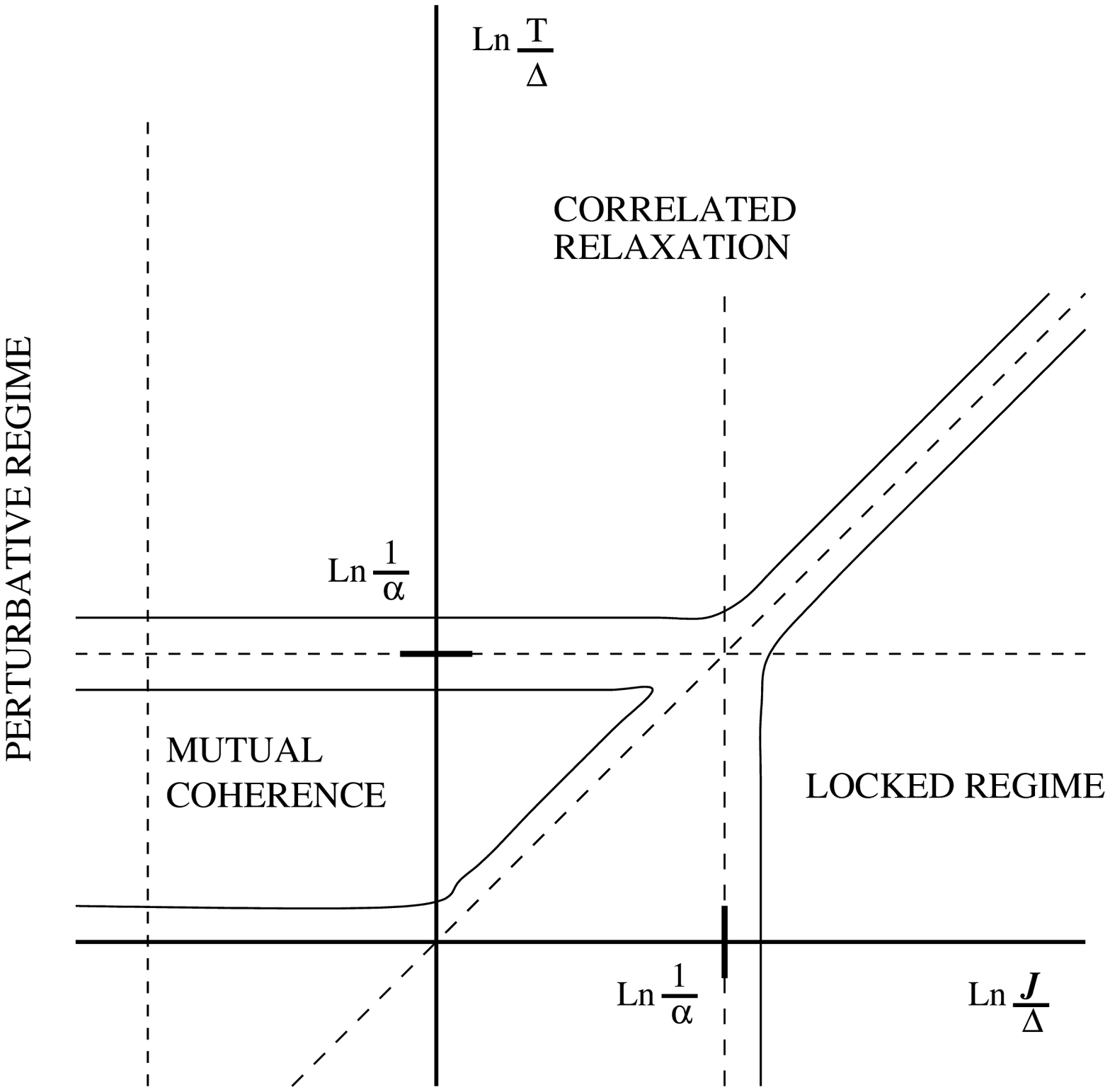}
\caption{}
\label{pphase}
\end{figure}

\section{The Measurement Operation}

We assume our 2 systems are coupled by a renormalised 
``ferromagnetic'' interaction
${\cal K}(R)$ and also that (i) $\Delta_A \gg \Delta_s$
and (ii) ${\cal K} > \xi_A \gg \Delta_{\beta}, kT$. Requirement (i) 
means the apparatus reacts quickly to any change in the system
state. Requirement (ii) means that the bias $\xi_A$ holds the 
apparatus in state
$|\Uparrow>$ when there is no coupling ; when 
${\cal K}$ acts, it must overcome this bias.  
The resulting net bias between initial and final states of the
apparatus must be $\gg \Delta_A, kT$, so that
apparatus transitions are irreversible.

We immediately see
that this system will behave according to eq. (\ref{2}).
The combined system-apparatus starts off either in $|\Uparrow \uparrow>$
(and stays there) or in $|\Uparrow \downarrow>$ (and then 
tunnels inelastically to $|\Downarrow \downarrow>$).
Typically the apparatus-bath coupling is
strong, so the apparatus relaxes quickly.  
Notice that the quantum dynamics of the {\it system} 
is {\it frozen} as soon as the apparatus-system coupling is switched on,
since ${\cal K} \gg \Delta_s$ the measurement
suppresses the quantum dynamics of the measured system.
This is a general feature of 
quantum measurements. Notice also the  
{\it dissipative} effect of the remaining oscillators (with 
$\omega_k < \Omega_o$) on the system is quite different once  
${\cal K}$ is switched on -- dissipative effects are stronger when a
system is biased. However the most interesting result comes from the 
mutual coherence regime- if accessible (eg., by varying $R$), we can set
up a situation where quantum interference {\it between system and apparatus}
can mess up the usual ``classical'' behaviour of the apparatus. In such a
situation, all the concepts of measurement theory, based on a classical
apparatus, would be incorrect. This 
result, if seen experimentally, would be intriguing at the very least
(as well as further testing quantum mechanics at the macroscopic scale).

Thus models of this kind, including apparatus
{\it and} system (as well as the environment), have interesting
light to shed on measurement theory.  The PISCES model appears to be the
first attempt to discuss all 3 partners in the measurement
operation on an equal and fully
quantum-mechanical level.

\end{document}